\documentclass[journal]{IEEEtran}
\usepackage{graphicx}
\usepackage{amsmath}
\usepackage{braket}
\usepackage{array}
\usepackage{makecell}
\usepackage{textcase} 
\begin{document}
\title{Traceable In Situ Microwave Power Measurement at the Cryogenic Device Plane in a Dilution Refrigerator}
\author{Paolo~Panetta\IEEEauthorrefmark{4},
Andrea~Celotto\IEEEauthorrefmark{1}\IEEEauthorrefmark{3},
Alessandro~Alocco\IEEEauthorrefmark{1}\IEEEauthorrefmark{3},
Bernardo~Galvano\IEEEauthorrefmark{2}\IEEEauthorrefmark{3},
Luca~Fasolo\IEEEauthorrefmark{3},
Emanuele~Palumbo\IEEEauthorrefmark{1}\IEEEauthorrefmark{3},
Luca~Callegaro\IEEEauthorrefmark{3},
Luca~Oberto\IEEEauthorrefmark{3},
Patrizia~Livreri\IEEEauthorrefmark{2},
and~Emanuele~Enrico\IEEEauthorrefmark{3}
\thanks{\IEEEauthorrefmark{1}Department of Applied Science and Technology, Politecnico di Torino, Torino, Italy.}
\thanks{\IEEEauthorrefmark{2}Department of Engineering, University of Palermo, Palermo, Italy.}
\thanks{\IEEEauthorrefmark{3}Istituto Nazionale di Ricerca Metrologica (INRiM), Torino, Italy.}
\thanks{\IEEEauthorrefmark{4}Department of Physics, Universit\`a degli Studi di Pisa, Pisa, Italy.}
\thanks{Corresponding author: Emanuele Enrico (e.enrico@inrim.it).}}

\maketitle

\begin{abstract}
Accurate knowledge of the microwave power delivered to a cryogenic device under test (DUT) is essential for the characterization and operation of superconducting quantum circuits. However, this information is difficult to obtain inside dilution refrigerators because of distributed attenuation, impedance mismatch, switch-path repeatability, and temperature-dependent microwave components. This paper presents an in situ measurement method for RF power at the cryogenic device plane. The method uses a custom variable temperature stage (VTS) as a cryogenic thermal-transfer element. The TVS is alternately heated by a four-wire DC heater and by microwave power dissipated in a 20 dB pass-through attenuator. By fitting the thermal transients and comparing the corresponding steady-state temperatures, the absorbed microwave power is inferred from a directly measured DC electrical power through an AC/DC substitution procedure. The finite reflection and transmission of the attenuator are then accounted for by cryogenic two-port scattering-parameter measurements based on a switch-assisted Short--Open--Load--Reciprocal calibration, so that the result is referred to the DUT reference plane. The system is demonstrated in a dilution refrigerator with powers between $-43$ and $-58$ dBm at the DUT input plane. The demonstrated relative standard uncertainty ranges from about $2\%$ at $43.9$ dBm to about $40\%$ at $57.6$ dBm. The proposed approach combines thermal RF power transfer, cryogenic S-parameter correction, and uncertainty evaluation in a measurement architecture compatible with quantum-device experiments, providing a practical route toward traceable microwave-power calibration at millikelvin stages.
\end{abstract}


\section{Introduction}
\IEEEPARstart{M}{icrowave} signals are central to the control, readout, and characterization of superconducting and other cryogenic quantum devices. In circuit quantum electrodynamics and related superconducting-qubit platforms, the electromagnetic field at the device reference plane determines transition rates, readout photon number, nonlinear operating points, measurement backaction, and unwanted effects such as ac Stark shifts or excess heating \cite{Devoret2013SuperconductingCircuits, Kjaergaard2020SuperconductingQubits, Blais2020NatureCircuitQED, Blais2021CircuitQED, Clerk2010QuantumNoise, Krantz2019QuantumEngineer,celotto2026deviceagnosticmicrowavenoisemetrology}. As these experiments are performed at millikelvin temperatures through heavily attenuated and thermally anchored coaxial lines, the power set at room temperature is non trivially related to the power delivered to the device under test (DUT). Cable losses, attenuator values, connector repeatability, switch paths, impedance mismatch, and temperature-dependent component behavior make the DUT-plane power a measurement quantity that must be calibrated rather than assumed.

Conventional RF and microwave power metrology relies extensively on thermal sensors and on DC-substitution or AC/DC transfer principles: the same thermal response is produced by an RF signal and by a measurable DC electrical power, thereby linking the RF power to voltage, current, and resistance standards \cite{Macpherson1955MicrowaveMicrocalorimeter, Larsen1977NBSTypeIV, Brunetti2008CoaxialMicrocalorimeter}. Extending this concept to cryogenic environments is attractive because it eliminates the uncertainty contributions added from routing the signal back to room temperature and calibrating the gain and noise of an output amplification chain. However, the cryogenic implementation is challenging. The power sensor must operate with a very small thermal load, remain compatible with the available cooling power and wiring, and provide an uncertainty model that includes both thermal and microwave effects.

Recent work has started to address this gap. Nanobolometers and graphene-based thermal detectors have demonstrated the relevance of calorimetric microwave detection for circuit-QED experiments \cite{Kokkoniemi2019Nanobolometer,Kokkoniemi2020BolometerThreshold}. A nanobolometer operated at low temperature has demonstrated broadband and traceable absorbed-power measurements based on DC substitution, including calibration of a heavily attenuated input line with uncertainty down to 0.1 dB at an absorbed power of about $-114$ dBm \cite{Girard2023CryogenicSensor}. A complementary approach has shown SI-traceable RF and microwave power measurements down to 3 K using a commercial thermoelectric power sensor, for RF levels from $-35$ dBm to 0 dBm over 100 kHz--10 GHz \cite{Celep2024SITraceableCryogenicRFPower}. A recent experiment demonstrated AC/DC power transfer by measuring the added noise of a cryogenic on-chip attenuator through the amplification chain \cite{descamps2026situcalibrationmicrowaveattenuation}. These results show the relevance of cryogenic RF power metrology. However, to the best of the authors' knowledge, a comprehensive analysis, supported by detailed uncertainty budgets, is still missing.

%they do not directly cover the operating regime targeted here: in situ measurement of the power available to a cryogenic DUT in a dilution refrigerator, around $-40$ to $-60$ dBm, with a device that can be inserted in the same microwave path used for DUT characterization.

A second, closely related requirement is to move the microwave reference plane to the DUT. VNA calibration and error-correction techniques are well established at room temperature \cite{Rumiantsev2008VNACalibration, Ferrero1992UnknownThru} and, in the last decades, have been extended and adapted for cryogenic temperatures as well
%At cryogenic temperatures, Ranzani \emph{et al.} implemented an adapted TRL calibration at the coldest stage of a dilution refrigerator, demonstrating two-port microwave calibration of common 50~$\Omega$ components at millikelvin temperatures \cite{Ranzani2013MilliKelvinTRL}. Subsequent work has extended the cryogenic calibration landscape to single-port resonator measurements, qubit drive-line components, full two-port S-parameter measurements for superconducting quantum integrated circuits, broadband multiline TRL coaxial calibration down to tens of millikelvin, and multiport connectorized devices at 3 K and above 
\cite{Ranzani2013MilliKelvinTRL,Simbierowicz2022QubitDriveLine,Shin2024BroadbandCryogenicTRL,Stanley2024TIMFourPortCryogenic,Stanley2025TIMAccuracyCryogenic}. In this work, the microwave correction of the VTS attenuator relies on a Short--Open--Load--Reciprocal (SOLR) calibration, i.e., an unknown-thru VNA calibration derived from the use of a reciprocal transmissive standard \cite{Ferrero1992UnknownThru}. The same cryogenic measurement architecture and its uncertainty treatment have been developed for full two-port calibrated S-parameter measurements at the millikelvin stage \cite{Oberto2025mKSParameterCalibration}.

The contribution of this paper is an instrumentation and measurement method that combines these two ingredients---AC/DC thermal power transfer and cryogenic S-parameter calibration---in a single dilution-refrigerator  to measure the incident RF power at the DUT input reference plane. A macroscopic copper variable temperature stage is weakly coupled to the refrigerator cold finger and equipped with a DC heater, a thermometer, and a microwave attenuator used as the RF absorbing element. By alternating DC and RF heating, fitting the thermal transients, and comparing the extracted steady-state temperatures, the microwave power dissipated in the attenuator is inferred from a measured DC power. The finite microwave absorption of the attenuator is then corrected using its calibrated scattering parameters, allowing to obtain the RF power at the DUT input reference plane. Following the approach of the Guite to the Expression of Uncertainty in Measurement (GUM) \cite{JCGM100_2008}, an uncertainty budget is developed by combining contributions associated with electrical-power measurement, thermal modelling, parasitic heating fluctuations, cold-stage temperature drift, and RF calibration.

This paper is organized as follows. Section II describes the measurement system, the modified VTS, thermometry, DC power measurement, and RF line calibration. Section III introduces the thermal model used to relate the DC and RF heating experiments. Section IV reports the measurement protocol. Section V presents the extracted RF power and the uncertainty contributions. Section VI concludes the paper and discusses the remaining limitations and possible improvements.

\section{Measurement system}
\begin{figure*}
    \centering
    \includegraphics[width = \textwidth]{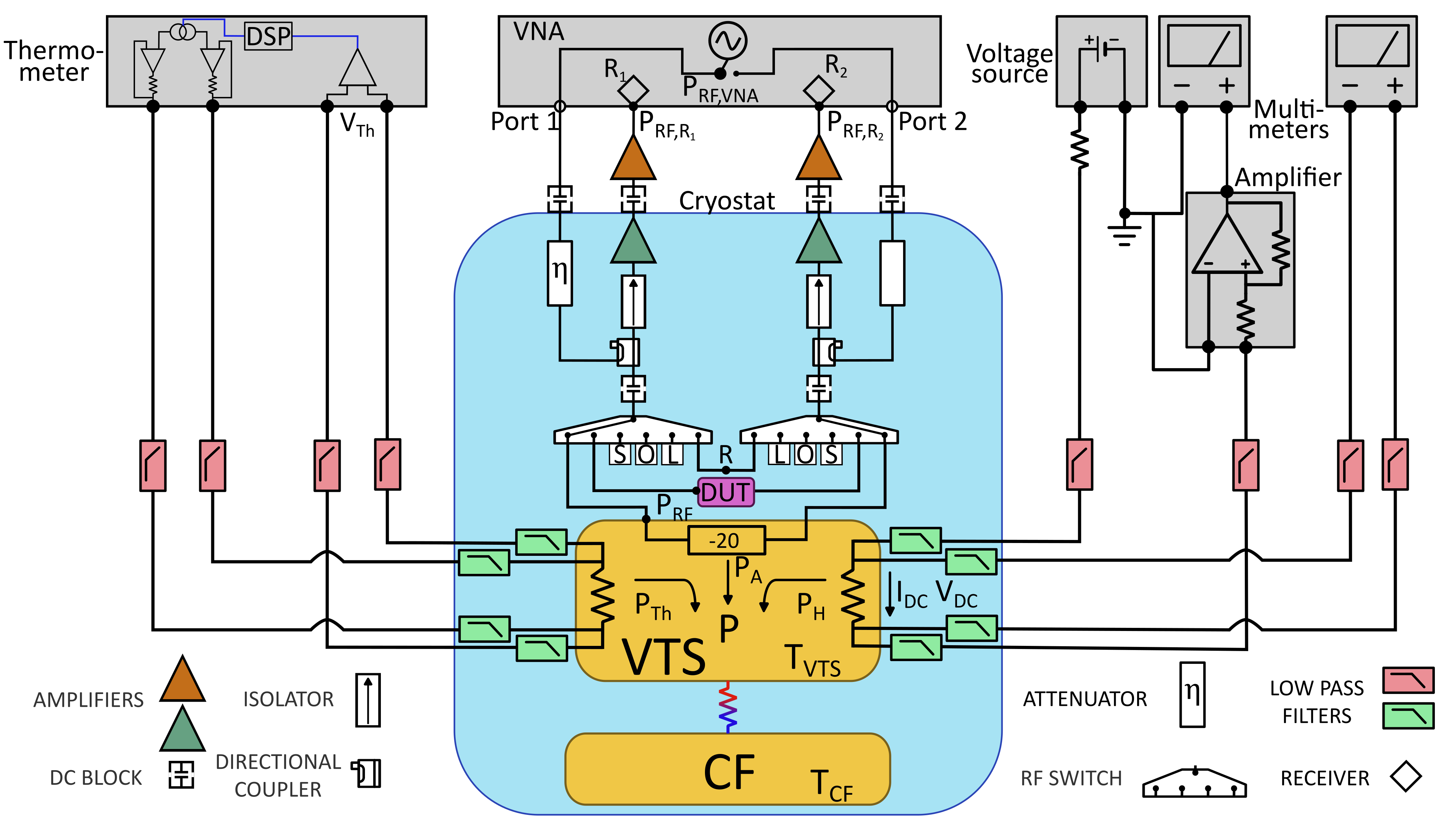}
    \caption{Schematic of the experimental setup. A Variable Temperature Stage (VTS) is weakly thermally linked with the Cold Finger (CF) of a dilution refrigerator with baseplate temperature of $\sim 80$ mK. VTS can be heated both with DC and RF through, respectively, a heater ($P_\mathrm{H}$) and a $20$ dB attenuator ($P_\mathrm{A}$). Its temperature is measured with a resistance thermometry bridge (Th). Both Th and H are driven in DC with a 4-wire configuration. Two nominally equal RF lines are driven by Port 1 and Port 2 and read by receivers $\mathrm{R_1}$ and $\mathrm{R_2}$ of a Vector Network Analyzer (VNA). The input lines are attenuated at multiple stages and we call $\eta$ the attenuation of the line connected to Port 1. The lines go through a directional coupler and a DC before finally reaching two 6-ports RF cryogenic switches. Output lines are isolated with an isolator and amplified in two different stages. DUT, VTS and a Reciprocal (R) are connected between the two switches. Short (S), Open (O) and Load (L) standards are also connected to the remaining ports of each switch. %The SOLR calibration is described in the text and in \cite{Oberto2025mKSParameterCalibration}.}
    }
    \label{fig:1}
\end{figure*}
\subsection{Overview}

A schematic representation of the experimental setup is shown in Fig.~\ref{fig:1}. A Variable Temperature Stage (VTS) is thermally weakly coupled to the Cold Finger (CF) of a dilution refrigerator. The VTS can be heated with both DC and RF through a heater and a $20$ dB attenuator, respectively. 
Two 6-ports RF switches are connected at the cold ends of four RF lines. The two input lines are attenuated at multiple stages with three $20$ dB attenuators \cite{Krinner2019}, while output lines are amplified with HEMTs inside the cryostat and LNAs at room temperature. On each switch are mounted Short (S), Open(O) and Load (L) impedance standards. The DUT, VTS and a Reciprocal (R) are connected between the two switches. % This scheme allows, through line calibration, to obtain the RF power delivered at DUT reference plane from the measure of the RF power dissipated in the VTS. 
The heater is driven in a 4 wire configuration, allowing the measurement of the current $I_\mathrm{{DC}}$ and voltage drop $V_\mathrm{DC}$ across the heater resistance. The VTS temperature is measured by means of a commercial resistance thermometry bridge. 
\subsection{Variable Temperature Stage}
The VTS is a custom object of macroscopic size ($\sim 50 \ \mathrm{cm}^3$) made entirely of OFHC Cu and plated in Au. It is attached to the CF through four joints of length $\sim 10$ cm. The joints are composed of external alumina washers with a stainless steel M2 screw inside, and are the dominant thermal link of the VTS to the CF, the other link being the superconducting cables used for electrical connections, as specified below. On VTS, a resistive heater H, a $20$ dB RF attenuator A and a RuOx thermistor Th are installed. Thermal contact between A and VTS, H and VTS, Th and VTS is a $\sim 1\ \mathrm{cm}^2$ interface between two Au-plated Cu planes. A power P is delivered to VTS through NbTiN superconducting cables and we name the power dissipated at heater, attenuator and thermometer $P_\mathrm{H}$, $P_\mathrm{A}$ and $P_\mathrm{Th}$, respectively. 
\begin{align*}
    P &= P_\mathrm{H}+P_\mathrm{A}+P_\mathrm{Th};\\
    P_\mathrm{H} &= P_\mathrm{DC,H} + P_\mathrm{N,H};  \\
    P_\mathrm{A} &= P_\mathrm{RF,A} + P_\mathrm{N,A};\\
    P_\mathrm{Th} &= P_\mathrm{exc,Th} + P_\mathrm{N,Th}, 
\end{align*}
where $P_\mathrm{DC,H}$ and $P_\mathrm{RF,A}$ are the signal DC and RF power at heater and attenuator, $P_\mathrm{exc,Th}$ is the excitation power of the thermometer and $P_\mathrm{N,H}$, $P_\mathrm{N,A}$ and $P_\mathrm{N,Th}$ are parasitic contributions to heating. Their nature and their effects on the measurement are treated in Section V.

\subsection{Measurement of dissipated power at heater}

The signal power at heater $P_\mathrm{DC,H}$ is measured by means of a 4-wire configuration (right side of Fig. \ref{fig:1}).
This bypasses the need for a calibrated cryogenic resistor and allows to measure $P_\mathrm{DC,H}$ as the product of the measured quantities $V_\mathrm{DC} $ and $I_\mathrm{DC}$.
The signal is filtered at room temperature (RT) with a cascaded RC filter and at cryostat baseplate with a cryogenic LC filter, allowing to minimize $P_\mathrm{N,H}$.

\subsection{Thermometry}
Measurements of $T_\mathrm{VTS}$ and $T_\mathrm{CF}$ are done with calibrated RuOx thermistors measured with a commercial multichannel resistance bridge (left side of Fig.~\ref{fig:1}). For visual clarity, only the thermistor on VTS and its wiring were drawn explicitly. For the CF, the configuration is the exact same as for VTS.  
%RF power dissipated at the attenuator $P_\mathrm{RF,A}$ is measured confronting heating due to a DC stimulus with heating due to a RF one. Under some assumption and approximations discussed in detail is section III, this can be done confronting steady state VTS temperatures when heated with DC or RF. These will depend indeed on heating power and CF temperature $T_\mathrm{CF}$.
As anticipated in Section I, the AC/DC transfer is performed by comparing the steady-state temperatures reached by the VTS under DC and RF heating. The thermometry chain is therefore used to extract reproducible thermal-equilibrium points rather than to measure the dissipated RF power directly.

\subsection{RF lines calibration}

Not all power $P_\mathrm{RF}$ delivered to the VTS port of the switch is dissipated in the attenuator. The latter quantity is related to the first through the VTS attenuator scattering parameters
\begin{equation}
    P_\mathrm{RF,A} = P_\mathrm{RF}\cdot(1-|S_{11}|_\mathrm{A}^2-|S_{21}|_\mathrm{A}^2).
    \label{VTS calibration}
\end{equation}
Since the quantity of interest is $P_\mathrm{RF}$, a S-parameter calibration is needed to measure $S_{11,\mathrm{A}}$ and $S_{21,\mathrm{A}}$ and invert \eqref{VTS calibration}.
%Several VNA calibration approaches can be used to move the measurement reference plane, including TRL and multiline TRL schemes and unknown-thru calibrations based on a reciprocal transmissive standard \cite{Rumiantsev2008VNACalibration,Engen1979TRL,Marks1991MultilineTRL,Ferrero1992UnknownThru}. In cryogenic environments, TRL-based calibrations have been implemented at millikelvin temperatures \cite{Ranzani2013MilliKelvinTRL,Shin2024BroadbandCryogenicTRL}, while SOLR-based approaches are attractive when compact coaxial standards and a reciprocal connection can be switched in during the same cooldown \cite{Ferrero1992UnknownThru,Oberto2025mKSParameterCalibration}.
We implemented the SOLR calibration in the same switch-based setup described in \cite{Oberto2025mKSParameterCalibration}, which also reports a complete uncertainty budget. Short (S), Open (O), and Load (L) impedance standards are connected to three ports of each switch, and the fourth standard is a Reciprocal (R), i.e., two cables nominally identical to the ones connecting the switches to the VTS and the DUT.
The calibration allows the full scattering matrix of any two-port device connected between the two switches, i.e., the VTS and the DUT in our case, to be de-embedded. Under the assumption, included in the calibration uncertainty budget, that the two reciprocal-path cables are identical, $P_\mathrm{RF}$ is equal to the power delivered at the DUT reference plane.

\section{Thermal model}

Since it is a bulk macroscopic object, electronic and phononic populations in VTS are assumed to be thermalized. Therefore one can describe the VTS as a thermodynamic object with a temperature $T_\mathrm{VTS}$ and a thermal capacitance $C_\mathrm{VTS} = C_\mathrm{VTS}(T_\mathrm{VTS})$. The VTS is weakly coupled to the CF, which acts as a reservoir, through a thermal conductance $G = G(T_\mathrm{VTS},T_\mathrm{CF})$. The VTS is also coupled with three distinct objects, A, H and Th, with a temperature-dependent thermal conductance, $\sigma_\mathrm{Cu-Cu}(T)$, assumed to be the same for all three. Powers $P_\mathrm{H}$, $P_\mathrm{A}$, and $P_\mathrm{Th}$ are dissipated, respectively, on each object. %From now on, we assume the thermodinamic behaviour of these three components is, at least qualitatively, the same, as both the wiring and the contact interface with VTS are the same. 
Under the assumption that $P_\mathrm{H}$, $P_\mathrm{A}$ and $P_\mathrm{Th}$ are stationary, one can write
\[
\left\{
\begin{aligned}
   &C_\mathrm{A}\dot T_\mathrm{A} = P_\mathrm{A}-\sigma_\mathrm{Cu-Cu}\cdot(T_\mathrm{A}-T_\mathrm{VTS})\\
   & C_\mathrm{H}\dot T_\mathrm{H} = P_\mathrm{H}-\sigma_\mathrm{Cu-Cu}\cdot(T_\mathrm{H}-T_\mathrm{VTS})\\
    &C_\mathrm{Th}\dot T_\mathrm{Th} = P_\mathrm{Th}-\sigma_\mathrm{Cu-Cu}\cdot(T_\mathrm{Th}-T_\mathrm{VTS})\\
    &C_\mathrm{VTS}\dot T_\mathrm{VTS} = \sigma_\mathrm{Cu-Cu}\cdot(T_\mathrm{A}+T_\mathrm{H}+T_\mathrm{Th}-\nonumber\\
    &\quad -3T_\mathrm{VTS})- G\cdot(T_\mathrm{VTS}-T_\mathrm{CF})
    \label{eq: sistema con A,H,Th}
\end{aligned}
\right.
\]
Solving this system accounting for the temperature dependencies of all thermal capacitance and thermal conductance is beyond the scope of this article. We limit here to note that, at steady state, a temperature difference $\Delta T_\mathrm{A(H,Th)}=T_\mathrm{A(H,Th)}-T_\mathrm{VTS}$ establishes, 
\begin{equation}
    \Delta T_\mathrm{A(H,Th)} = \frac{P_\mathrm{A(H,Th)}}{\sigma_\mathrm{Cu-Cu}}.
\end{equation}
Thermal conductance of bolted Cu--Cu and Au-plated Cu--Cu interfaces at sub-kelvin and few-kelvin temperatures have been measured and reviewed in the literature \cite{Didschuns2004BoltedCopperJoints,Dhuley2019PressedCopperContacts,Ekin2006LowTemperatureMeasurements}. Using representative conductance values $>10^{-2}\ \mathrm{W\ K^{-1}}$ (Fig. 6 of \cite{Didschuns2004BoltedCopperJoints}) in the temperature range of interest and the experimental condition $P_\mathrm{A(H,Th)}<10^{-7}$ W gives $\Delta T_\mathrm{A(H,Th)} < 0.01$ mK. 
This value is well below the thermometric uncertainty reported in Section V and we can therefore assume $T_\mathrm{A(H,Th)}$ to be equal to $T_\mathrm{VTS}$ and write a single thermodynamic equation for VTS
%We therefore can neglect thermal exchange between VTS and A,H and Th and assume all the power is dissipated directly on the VTS. This leads to the following 
\begin{multline}
    C_\mathrm{VTS}\dot T_\mathrm{VTS}
    = P - G\cdot(T_\mathrm{VTS}-T_\mathrm{CF}) =\\ 
     = P_0 + P_\mathrm{DC,H} + P_\mathrm{RF,A} - G\cdot(T_\mathrm{VTS}-T_\mathrm{CF})
     \label{modello termodinamico}
\end{multline}
where $P_0 = P_\mathrm{N,H} + P_\mathrm{N,A} + P_\mathrm{N,Th} + P_\mathrm{exc,Th}$ is assumed constant throughout the whole experiment. As for $G$, it is the thermal conductance of the steel screws and alumina washers described in Section II-B. It is dominated by the low-temperature thermal conductivity of stainless steel, which is approximately linear in temperature over the range of this experiment %which can be approximated as $\sigma=\sigma_0T$ over the temperature range considered here 
\cite{Woodcraft2009LowTemperatureThermalConductivity,Ekin2006LowTemperatureMeasurements}. For a rod model of length $L$ and cross-section $A$ with fixed boundary conditions $T(0) = T_\mathrm{VTS}$ and $T(\mathrm{L}) = T_\mathrm{CF}$, the total conductance of the rod is $G = G_0\cdot (T_\mathrm{VTS} + T_\mathrm{CF})/2$ with $G_0 = \sigma_0\cdot\mathrm{A/L}$. The adequacy of this frist-order conductance model over the investigated temperature range is assessed a posteriori from the DC-heating calibration curve and its residuals (Fig. \ref{fig:3}).
$G$ can be plugged into \eqref{modello termodinamico} and analytical solutions exist for $T_\mathrm{VTS}(t)$ of the form
\begin{equation}
    T_\mathrm{VTS}(t) = T_\mathrm{fin,VTS}\frac{1-Be^{-2t/\tau}}{1+Be^{-2t/\tau}}
    \label{soluzione dinamica}
\end{equation}
here $T_\mathrm{fin,VTS}$ is the steady state temperature of VTS. These are easily calculated from \eqref{modello termodinamico} and their values are
\begin{equation}
    T_\mathrm{fin,VTS}^\mathrm{DC} = \sqrt{T_\mathrm{CF}^2 + 2\frac{P_0}{G_0}+ 2\frac{P_\mathrm{DC,H}}{G_0}}
    \label{steady state DC}
\end{equation}
\begin{equation}
    T_\mathrm{fin,VTS}^\mathrm{RF} = \sqrt{T_\mathrm{CF}^2 + 2\frac{P_0}{G_0}+ 2\frac{P_\mathrm{RF,A}}{G_0}}
    \label{steady state RF}
\end{equation}
where superscript indicates whether that $T_\mathrm{fin,VTS}$ was obtained heating with DC or RF. 

Equations \eqref{steady state DC} and \eqref{steady state RF} allow, under hypothesis of stationarity of $P_0$ over the whole experiment, to do the AC/DC power transfer by fitting \eqref{steady state DC} over an arbitrary stepped range of $P_\mathrm{DC,H}$ and then inverting \eqref{steady state RF} to obtain $P_\mathrm{RF,A}(T_\mathrm{fin,VTS}^\mathrm{RF})$ as 
\begin{equation}
    P_\mathrm{RF,A} =  G_0 \frac{(T_\mathrm{fin,VTS}^\mathrm{RF})^2-T_\mathrm{CF}^2}{2} - P_0
    \label{steady state inverted}
\end{equation}
Note that here stationarity is required only for $P_0$ over the whole experiment, and $P_\mathrm{DC,H}$ and $P_\mathrm{RF,A}$ are varied with a stepwise sweep. For the dynamic to obey \eqref{soluzione dinamica}, though, $P_\mathrm{DC,H}$ or $P_\mathrm{RF,A}$ need to remain constant over the single power step (see Fig.~\ref{fig:2}(a)).   
One can then correct for VTS S parameters in order to extract $P_\mathrm{RF}$ using equation \eqref{VTS calibration}.

\section{Measurement protocol}
Before cooldown, a preliminary characterization of the RF line was done to extract a rough estimate of the total attenuation from source to DUT reference plane $\eta' \simeq 57$ dB. Once cooled, VTS reached a base temperature with no external heating of $228$ mK, measured to be stable within $1$ mK over a five hours time span. 

VTS was then heated alternately with DC and RF signal at $4$ GHz over a 16 dB span between $-43$ dBm and $-58$ dBm, with a power step of $4$ dB and a step duration $\Delta t = 77$ minutes. RF source power $P_\mathrm{RF,VNA}$ was increased of $\eta'$ with respect to the desired power at DUT reference plane. When heated in DC, polarity of signal was switched every $\sim 5$ minutes (see inset of Fig.~\ref{fig:2}(a)). This allows to remove, during analysis, current or voltage biases. $I_\mathrm{DC}$, $V_\mathrm{DC}$, $T_\mathrm{VTS}$ and $T_\mathrm{CF}$ were measured every $10$ seconds, to allow a long integration time. 

Finally, the SOLR calibration procedure described in \cite{Oberto2025mKSParameterCalibration} was applied to measure the VTS scattering parameters at the switch-defined reference planes. These calibrated values were then used to evaluate the absorbed-power correction in \eqref{VTS calibration}.

All drive and measurement instruments were simultaneously operated by a dedicated software.

\section{Results}
\subsection{Steady-state temperature extrapolation and uncertainty evaluation}
\begin{figure}
    \centering
    \includegraphics[width = \linewidth]{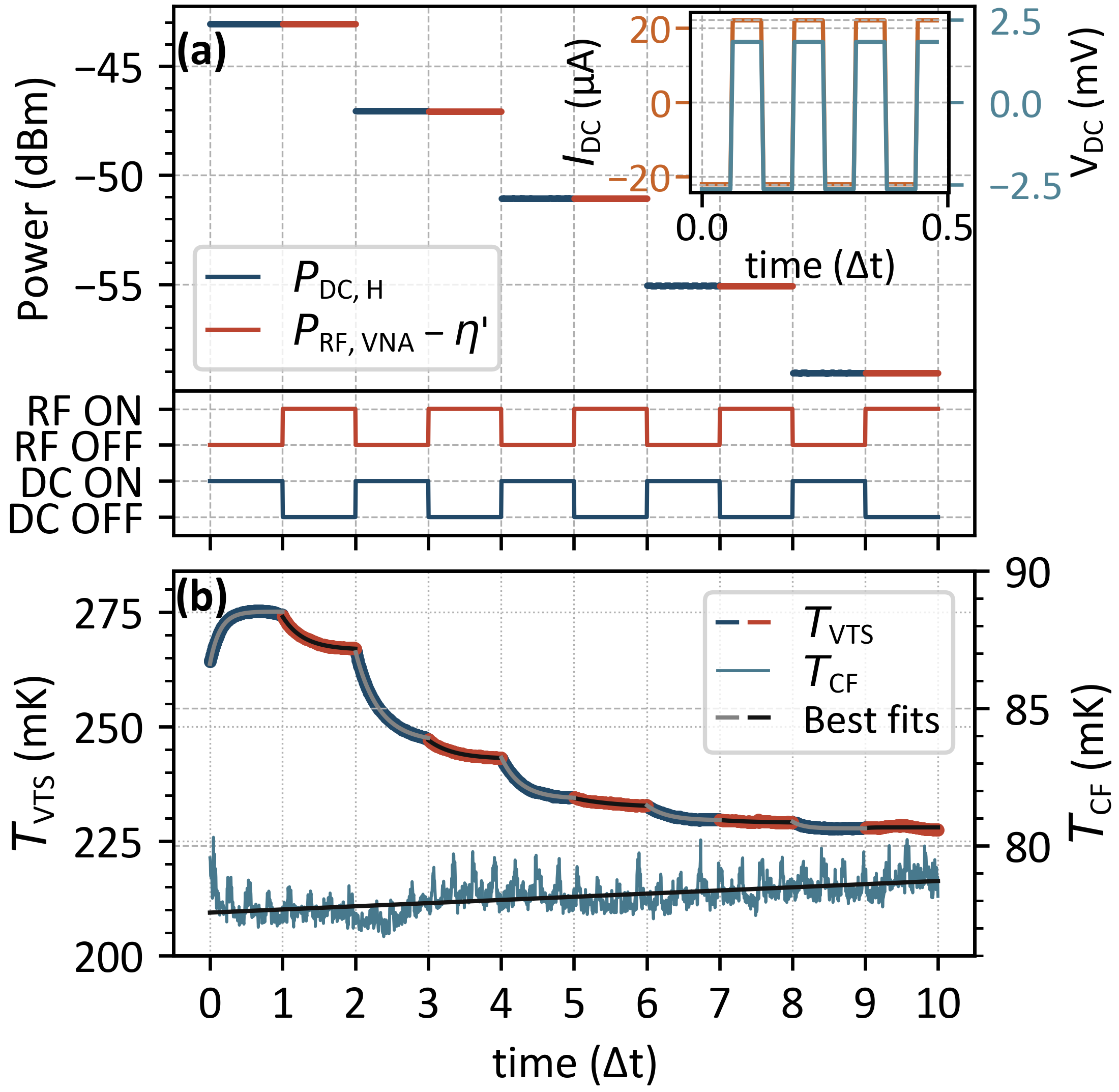}
    \caption{Here, the time axis is normalized for $\Delta t = 77$ minutes.  (a) Power vs time plots. $P_\mathrm{DC,H}$ is the measured power dissipated at H. $P_\mathrm{RF,VNA}$ is the power delivered at VNA source and  $\eta'$ is a preliminarily estimated measure of the total attenuation between source and DUT reference plane. VTS was alternately heated with DC and RF drives with a power step of $4$ dB. When driven in DC, polarity of signal was swapped every $\sim 5$  minutes to account for current amplifier drifts. In the inset, measured $V_\mathrm{DC}$ and $I_\mathrm{DC}$ are plotted.
    (b) $T_\mathrm{VTS}$ and $T_\mathrm{CF}$ vs time plots. $T_\mathrm{CF}$ was measured to slightly drift of $0.09$ mK/hour. Each interval of length $\Delta$t of $T_\mathrm{VTS}$ was fit with \eqref{soluzione dinamica}.}
    \label{fig:2}
\end{figure}
Values of $T_\mathrm{fin,VTS}^\mathrm{DC}$ and $T_\mathrm{fin,VTS}^\mathrm{RF}$ were obtained by fitting $T_\mathrm{VTS}(t)$ with \eqref{soluzione dinamica} for each time interval $\Delta t$ and keeping $T_\mathrm{fin,VTS}$, $B$ and $\tau$ as fit parameters. The superscript DC or RF indicates whether the value of $T_\mathrm{fin,VTS}$ was obtained by heating with DC or RF. Fitting $T_\mathrm{fin,VTS}$ instead of simply waiting for $T_\mathrm{VTS}$ to stabilize allows to find its central value even if $\Delta t\gg\tau$ does not hold. Data and best fit curves are reported in Fig. \ref{fig:2}(b). For visual clarity a color code is used for $T_\mathrm{VTS}$, assigning blue to DC heating and red to RF heating. Best fits are plotted in gray over the data. 

The uncertainty associated with $T_\mathrm{fin,VTS}$ includes three contributions: the uncertainty of the transient fit, the stability of the CF temperature $T_\mathrm{CF}$, and the fluctuations of the parasitic background power $P_0$. The corresponding budget is summarized in Table \ref{tab: T_fin_VTS} for $T_\mathrm{fin,VTS} = 229.6$ mK.

\begin{table}
    \centering
    \caption{Uncertainty budget on $T_\mathrm{fin,VTS} = 229.6$ \textnormal{mK}}
    \vspace{-3 mm}
    \begin{tabular}{lcc}
        \hline
        \hline
        Uncertainty source & \makecell{Uncertainty \\contribution / mK} \\%\makecell{Uncertainty \\percentage} \\
        \hline
         Fit uncertainty & $0.07$ \\ %& 0.2 - 13.1 \\
         $T_\mathrm{CF}$ stability & $0.23$\\ %& 9.6 - 13.4  \\
         Fluctuations of $P_0$ & $0.58$ \\% & 75.5 - 90.0\\
         \hline
         Total uncertainty & 0.63 \\%& 100\\
         \hline
         \hline
         
    \end{tabular}
    \label{tab: T_fin_VTS}
\end{table}
The contribution associated with the CF drift was evaluated from the linear trend observed in $T_\mathrm{CF}$ during the experiment (lower curve in Fig. \ref{fig:2}(b)). An average drift of $0.09$ mK/hour was measured, for a total drift of $1.15$ mK over the whole experiment. Since the VTS has a large thermal capacitance and therefore averages fast temperature fluctuations, using the standard deviation of the full $T_\mathrm{CF}$ time series would overestimate its effect on $T_\mathrm{fin,VTS}$. The drift contribution was therefore estimated from the measured drift coefficient multiplied by the experiment duration, assuming a rectangular distribution. Propagation through \eqref{steady state DC} and \eqref{steady state RF} gives a contribution between 0.03 mK and 0.05 mK, depending on the value of $T_\mathrm{fin,VTS}$.

The dominant contribution arises from slow fluctuations of the parasitic background power $P_0$. This term includes the thermometer drive power and residual parasitic heating mechanisms affecting the heater, attenuator, thermometer, and wiring. Possible sources include heat conduction through cable shields, radiation from warmer stages, electrical noise, parasitic currents, ground loops, and signal dissipation in cables. The thermometer drive power is expected to be stationary and well below 1 nW, while the remaining contributions are more difficult to model from frist principles.
To evaluate their effect experimentally, a five-hour measurement was performed with no applied DC or RF signal power, i.e. $P_\mathrm{DC,H} = P_\mathrm{RF,A} = 0$. The observed maximum excursion of $T_\mathrm{VTS}$ was 1.06 mK. Since the contribution from $T_\mathrm{CF}$ drift over this time scale is only a few $10^{-2}$ mK, the observed variation was assigned to fluctuations of $P_0$ and treated as a rectangular distribution. As shown in Table \ref{tab: T_fin_VTS}, this term dominates the uncertainty of $T_\mathrm{fin,VTS}$, contributing between $75.5\%$ and $90\%$ of the total variance. The resulting standard uncertainty on $T_\mathrm{fin,VTS}$ lies between $0.61$ mK and $0.66$ mK.

\subsection{DC-to-thermal calibration}
the DC-heating data were used to calibrate the thermal response of the VTS. 
An Orthogonal Distance Regression (ODR) \cite{Boggs1990ODR} fit was performed on 
$T_\mathrm{fin,VTS}^\mathrm{DC}(P_\mathrm{DC,H})$ with \eqref{steady state DC} (blue data in Fig. \ref{fig:3}). It minimizes the normalized orthogonal distance between data and best fit curve, accounting for both uncertainty on $P_\mathrm{DC,H}$ and on $T_\mathrm{fin,VTS}^\mathrm{DC}$. The uncertainty on $P_\mathrm{DC,H}$ was evaluated as in Table \ref{tab: P_DC} from uncertainties on measured voltage $V_\mathrm{DC,H}$ and current $I_\mathrm{DC,H}$. The cold finger temperature was treated as a fixed parameter and set equal to the mean value $\bar T_\mathrm{CF}$ measured over the whole experiment. The best fit parameters were
\begin{equation*}
    G_0 = (4.03 \pm 0.07) \cdot 10^{-6}\mathrm{\ W\ K^{-2}}
\end{equation*}
and 
\begin{equation*}
    P_0 = 90.7 \pm 1.8\ \mathrm{nW}.
\end{equation*}

The extracted value of $P_0$ is in agreement with Table 2 of \cite{Krinner2019}, where an heat load between $13$ nW and $30$ nW (depending on cable specifics) per line was measured on the MXC.

The extracted value of $G_0$ is consistent with an independent estimate based on the geometry of the four stainless-steel screws and on literature values of the low-temperature thermal conductivity of stainless steel. Using $G_{0,\mathrm{est}} = 4\sigma_0\mathrm{A}/\mathrm{L}$, with the factor 4 accounting for the four nominally identical screws, A = $0.64\ \pi \ \mathrm{mm^2}$, L = $10$ cm and $\sigma_0 = (0.5\div1)\cdot10^{-3}\ \mathrm{W\ cm^{-1}K^{-2}}$ (Fig. 3.21 of \cite{Pobell2007}), one finds $G_{0,\mathrm{est}} \simeq (4\div8)\cdot10^{-6}\mathrm{WK^{-2}}$, in agreement with the fitted value. Moreover, the residuals of the fit, shown in the lower panel of Fig. \ref{fig:3}, do not exhibit a systematic dependence on power. These two observation provide an internal consistency check of the first-order conductance model used for deriving \eqref{soluzione dinamica}.

% An  Orthogonal Distance Regression (ODR) \cite{Boggs1990ODR} fit was performed on $T_\mathrm{fin,VTS}^\mathrm{DC}(P_\mathrm{DC,H})$ with \eqref{steady state DC} (blue data in Fig.~\ref{fig:3}), keeping $T_\mathrm{CF}$ as a fixed parameter with value $\bar T_\mathrm{CF}$, calculated as the mean CF temperature over the whole experiment. Uncertainties on $P_\mathrm{DC,H}$ were calculated from uncertainties on $I_\mathrm{DC}$ and $V_\mathrm{DC}$. Table \ref{tab: P_DC} reports the uncertainty budget for $P_\mathrm{DC,H} = -59.06$ dBm. The uncertainty budget on $T_\mathrm{fin,VTS}$ is schematized in Table \ref{tab: T_fin_VTS} and accounts for the drift of $T_\mathrm{CF}$, fluctuations of $P_0$ and best fit parameters uncertainties on \eqref{soluzione dinamica}.\\
% Finally, \eqref{VTS calibration} and \eqref{steady state inverted} were used to obtain $P_\mathrm{RF}$ and uncertainty budgets (Table \ref{tab: P_RF - 57 dBm} and Table \ref{tab: P_RF -43 dBm}) were evaluated. They take into account error propagation from \eqref{VTS calibration} and \eqref{steady state inverted} and corrects for a systematic error due to cable heating.  

\begin{table}
    \centering
    \caption{Uncertainty budget on $P_\mathrm{DC,H} = 1.2407$ \textnormal{nW} ($-59.06$ \textnormal{dBm})}
    \vspace{-3 mm}
    \begin{tabular}{lcc}
        \hline
        \hline
        Uncertainty source & \makecell{Uncertainty \\contribution / fW}\\
        %& \makecell{Uncertainty \\ percentage}\\
        \hline
         Type A uncertainty & 0.2 \\
         Transimpedance amplifier & 7.2 \\
         Voltage measurement  & 6.2 \\
         Current measurement & 0.1 \\
         \hline
         Total uncertainty & 9.5\\
         \hline
         \hline
         
    \end{tabular}
    \label{tab: P_DC}
\end{table}

\subsection{RF power extraction at the attenuator}

\begin{figure}
    \centering
    \includegraphics[width = \linewidth]{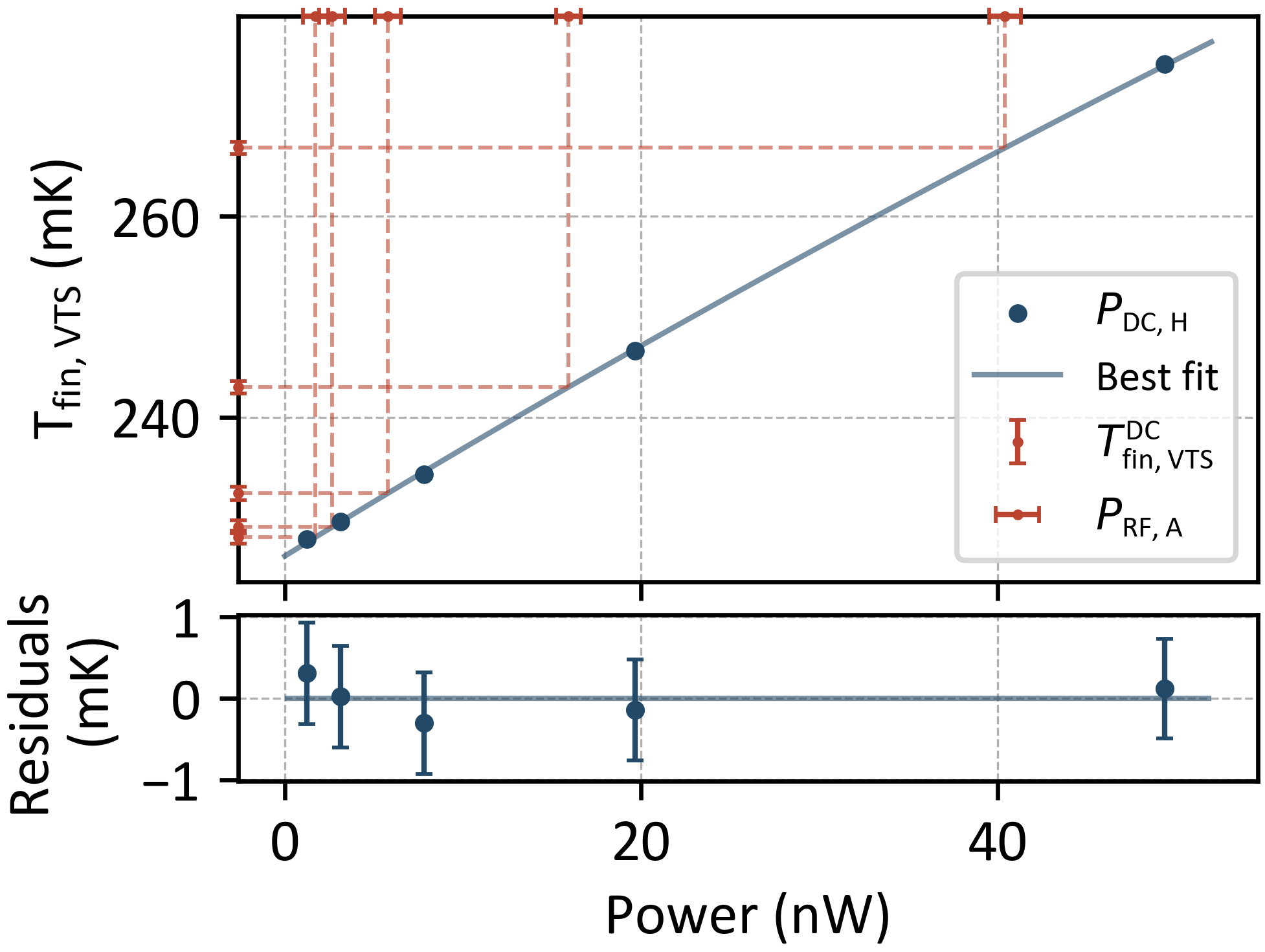}
    \caption{$T_\mathrm{fin,VTS}$ vs Power. $P_\mathrm{DC,H}$ data were fit with \eqref{steady state DC}. Then \eqref{steady state RF} was inverted to obtain $P_\mathrm{RF,A}(T_\mathrm{fin,VTS})$. A visual representation of the inversion idea is graphed in red. $T_\mathrm{fin,VTS}^\mathrm{RF}$ on the y-axis are projected onto the best fit curve and then projected again on the x-axis. Uncertainties on the best fit parameters are taken into account by the uncertainty propagation formula of \eqref{steady state inverted}. Below, residuals of $P_\mathrm{DC,H}$ from the best fit curve.}
    \label{fig:3}
\end{figure}
The microwave power $P_\mathrm{RF,A}$ dissipated in the VTS was then extrapolated using \eqref{steady state inverted}. It is the inverse of \eqref{steady state RF}, obtained by solving for $P_\mathrm{RF,A}$ and using for $P_0$ and $G_0$ the best fit values on the DC data.
The uncertainty on $P_\mathrm{RF,A}$ was obtained by propagating the uncertainty of $T_\mathrm{fin,VTS}$, $T_\mathrm{CF}$, $G_0$, $P_0$ and the covariance between $G_0$ and $P_0$. A graphic representation of the calculation is presented in Fig. \ref{fig:3}, where the values of $T_\mathrm{fin,VTS}^\mathrm{RF}$ (red errorbars on the y-axis) are projected onto the best fit curve and then projected again onto the x-axis.

A correction was also applied for the RF power dissipated in the superconducting cable section connected to the VTS and heating the attenuator. The DC contribution is negligible because the relevant cables are NbTiN superconducting cables. For the RF signal, the dissipated heat can be estimated as the product of the cable attenuation coefficient with the cable length and the RF power. Using a typical attenuation coefficient of about $0.3 $ dB/m below $4 $ K and a cable lenght of approximately $15$ cm gives a dissipated fraction of about $1\%$ of the transported RF power. Assuming that half of this heat flows to the VTS, $P_\mathrm{RF,A}$ would be overestimated by approximately $0.5\%$. This correction was applied, and its uncertainty was represented by a rectangular distribution between $0\%$ and $1\%$.

\subsection{Correction to the DUT reference plane}
\begin{table}
    \centering
    \caption{Uncertainty budget on $P_\mathrm{RF}$ for $P_\mathrm{RF} = 1.74 \ \mathrm{nW}$  ($-57.58$ \textnormal{dBm})}
    \vspace{-3 mm}
    \begin{tabular}{lcc}
        \hline
        \hline
        Uncertainty source & \makecell{Uncertainty \\contribution / nW} & \makecell{Uncertainty \\percentage} \\
        \hline
        $G_0$, $P_0$, Cov($G_0$, $P_0$)  & $0.34$  & $24$\\
        $T_\mathrm{fin,VTS}^\mathrm{RF}$ & $0.58$ & $67$\\
        $T_\mathrm{CF}$ & $0.21$ & $9$ \\
        RF line calibration & $0.003$ & $<0.01$ \\
        Attenuation in cables & $0.005$ & $<0.01$\\
        \hline
        Total uncertainty & $0.71$ & $100$ \\
        \hline
        \hline
         
    \end{tabular}
    \label{tab: P_RF - 57 dBm}
\end{table}

\begin{table}
    \centering
    \caption{Uncertainty budget on $P_\mathrm{RF}$ for $P_\mathrm{RF} = 40.9 \  \mathrm{nW}$ ($-43.87$ \textnormal{dBm})}
    \vspace{-3 mm}
    \begin{tabular}{lcc}
        \hline
        \hline
        Uncertainty source & \makecell{Uncertainty \\contribution / nW} & \makecell{Uncertainty \\percentage} \\
        \hline
        $G_0$, $P_0$, Cov($G_0$, $P_0$)  & $0.55$  &  $37$\\
        $T_\mathrm{fin,VTS}^\mathrm{RF}$ & $0.67$ &  $55$\\
        $T_\mathrm{CF}$ & $0.21$ & $5$ \\
        VTS calibration & $0.08$ & $\sim 1$ \\
        Attenuation in cables & $0.12$ &  $\sim 2$\\
        \hline
        Total uncertainty & $0.90$ & $100$ \\
        \hline
        \hline
         
    \end{tabular}
    \label{tab: P_RF -43 dBm}
\end{table}

The quantity $P_\mathrm{RF,A}$ represents the RF power absorbed by the $20$ dB attenuator mounted on VTS. The target measurand, however, is the RF power available at the DUT reference plane. This is obtained by accounting for the finite reflection and transmission coefficients of the attenuator, as in \eqref{VTS calibration}. The scattering parameters of the attenuator were obtained from the cryogenic SOLR calibration described in Section II.E and in \cite{Oberto2025mKSParameterCalibration}. The uncertainty associated with this RF correction was then propagated into the final uncertainty budget.

The values of $P_\mathrm{RF}$ so extracted lie between $-43.87$ dBm and $- 57.58$ dBm with relative uncertainty going from $2\%$ for $P_\mathrm{RF}= -43.87$ dBm to $40\%$ for $P_\mathrm{RF} = -57.58$ dBm. As can be seen from Tables \ref{tab: P_RF - 57 dBm} and \ref{tab: P_RF -43 dBm}, the dominant contributions are the uncertainties on $T_\mathrm{fin,VTS}^\mathrm{RF}$ and on the best fit parameters $G_0$ and $P_0$, while the CF temperature drift plays a minor role accounting for less than $10\%$ of the total uncertainties. The uncertainties introduced by cables heating and RF lines calibration are orders of magnitude smaller than the latter and are therefore negligible. 

These results can be interpreted in this way:
\begin{itemize}
    \item for small signal powers the $T_\mathrm{fin,VTS}$ variations become too small with respect to their uncertainties; 
    \item the lowest signal power used for this experiment is $-57.58$ dBm, which corresponds to $1.75$ nW. This is approximately equal to uncertainty on $P_0$, indicating that for such low powers the signal becomes indistinguishable from the power noise. 
\end{itemize} 
 
If one considers that also the uncertainty on $T_\mathrm{fin,VTS}$ is dominated from fluctuations of $P_0$, it becomes clear that this is the first issue to be addressed in order to achieve sensitivity to lower energies.

\section{Conclusions}
An in situ microwave power measurement method for dilution-refrigerator experiments was presented. The method used a modified VTS as a cryogenic thermal transfer standard and realized an AC/DC comparison between microwave heating in a $20$ dB attenuator and directly measured DC heating in a four-wire resistor. By fitting the thermal relaxation of the weakly coupled VTS and comparing the corresponding steady-state temperatures, the RF power dissipated in the absorber was inferred from the DC electrical power. A cryogenic SOLR calibration was then used to correct for the finite reflection and transmission of the pass-through attenuator and to refer the result to the DUT reference plane.

The experiment demonstrated the feasibility of measuring RF powers in the  $-43$ to $-58$ dBm range inside a dilution refrigerator using instrumentation that is compatible with cryogenic quantum-device measurements. The approach is complementary to on-chip cryogenic bolometers and commercial cryogenic thermoelectric sensors: it targets a higher-power, DUT-plane calibration regime while preserving a direct link to electrical quantities through the DC substitution step. The pass-through implementation also allowed the calibrated power path to remain close to the one used during device characterization.

The present uncertainty budgets identify the dominant contributions as the stability of the power background $P_0$, the extraction of the VTS steady-state temperature and, to a minor extent, the stability of the thermal reservoir $T_\mathrm{CF}$. %the stability of the thermal background, the extraction of the VTS steady-state temperature, the DC power measurement, and the microwave correction associated with the attenuator scattering parameters and switch-based calibration. 
These results indicate that further improvements should focus on reducing parasitic heating and thermal drifts, as well as improving electrical filtering and shielding of the thermometer and heater lines. %and extending the RF calibration uncertainty analysis over the full operating band. 
A smaller VTS would reduce the response time scales, making the experiment faster and less sensitive to slow dynamics. This would also allow for an improved measurement protocol, in which a PID control system could perform the AC/DC transfer while keeping $T_\mathrm{VTS}$ fixed. 

Future work will also compare the VTS-based result with an independent cryogenic power standard and assess the method during operation with representative quantum devices. Overall, the proposed VTS-based power meter provides a practical route toward traceable RF power calibration at the cryogenic DUT plane, addressing a measurement need that is increasingly important for scalable quantum-device characterization.
\section{*Aknowledgments}
This work is supported by the European projects MiSS and MetSuperQ. MiSS is funded by the European Union through the Horizon Europe 2021-2027 Framework Programme, Grant agreement ID: 101135868. The 23FUN08 MetSuperQ project has received funding from the European Partnership on Metrology, co-financed from the European Union’s Horizon Europe Research and Innovation Programme and by the Participating States.
\bibliographystyle{IEEEtran}
\bibliography{references}

\end{document}